# Viscous Dissipation, Helicity and Enstrophy of Bouncing Droplets undergoing Off-center Collision

Chengming He, Xi Xia, Peng Zhang[*]

Department of Mechanical Engineering

The Hong Kong Polytechnic University, Hung Hom, Kowloon, Hong Kong

[*] Corresponding author
E-mail address: pengzhang.zhang@polyu.edu.hk (P. Zhang)

**Abstract**

The off-center collision of binary bouncing droplets of equal size was studied numerically by a volume-of-fluid (VOF) method with two marker functions, which has been validated by comparing with available experimental results. A non-monotonic kinetic energy recovery with varying impact parameters was found based on the energy budget analysis. This can be explained by the prolonged entanglement time and the enhanced internal-flow-induced viscous dissipation for bouncing droplets at intermediate impact parameters, compared with those at smaller or larger impact parameters. The universality of this non-monotonicity was numerically verified, and thereby an approximate fitting formula was proposed to correlate the kinetic energy dissipation factor with the impact parameter for various Weber numbers and Ohnesorge numbers. From the vortex dynamics perspective, a helicity analysis of droplet internal flow identifies a strong three-dimensional interaction between the "ring-shaped" vortices and the "line-shaped" shear layers for off-center collisions. Furthermore, we demonstrated theoretically and verified numerically that the equivalence between the total enstrophy and the total viscous dissipation, which holds for a single-phase flow system confined by stationary boundaries, is not generally satisfied for the two-phase flow system containing gas-liquid interfaces. This is attributed to the work done by the unbalanced viscous stresses, which results from the interfacial flow and the vorticity associated with the movement of the oscillating interface.

# I. INTRODUCTION

Binary droplet collision [1-8] had been studied substantially to probe more complicated physics in the context of various spray processes, such as cloud and raindrop formation [9,10], dense spray combustion [11,12], droplet manipulation in microfluidics [13,14], and various applications involving functional interfaces [15-20] in material science.

The investigation of droplet collision could date back to the study on cloud aerosol, which was aimed to explain the mechanism of raindrop formation. Historically, studying the raindrop formation raised some arguments about which mechanism plays the dominant role [9,10], between the fragmentation of large droplets and the coalescence of minute droplets. It then spawned the early studies [2,3,8] on the collision between two water droplets in standard atmosphere environment, resulting in the discovery of two collision outcomes, coalescence and separation. However, raindrops might have opposite charges [21] so that they tend to bounce off upon collision and their distributions of size, number density, and velocity are accordingly influenced.

In recent years, to understand the sprays of liquid fuels in combustion conditions, research efforts have been devoted to the collision between two hydrocarbon droplets [4,6] in various gaseous environments, and most relevant studies were focused on droplet coalescence [1,3,22] and the subsequent internal mixing [23-27]. Although not being sufficiently studied, interesting phenomena were also observed for droplet bouncing; for example, bouncing was found to occur to only fuel droplets [4,6] but not water droplets in standard atmosphere environment. Furthermore, the collision outcomes can be significantly affected by the gas environment; specifically, increasing the gas pressure promotes droplet bouncing and decreasing the gas pressure promotes droplet coalescence [6,28,29]. Given the elevated pressure [6,28] in real combustion chambers, droplet bouncing is a prominent collision outcome and of great significance in dense spray combustion. It has been verified both experimentally and numerically [30,31] that high pressure environment favors droplet bouncing in impinging jets.

Compared with the extensively studied droplet coalescence and separation [1,2,4,6,8,22,26], binary droplet bouncing has been investigated by only a few studies. Pan *et al.* [5] studied the evolution of energy budget for head-on bouncing between binary droplets of equal size both experimentally and numerically. Tang *et al.* [7] experimentally observed the head-on bouncing between binary droplets of unequal sizes. Recently, Zhang and Zhang [32] numerically studied the kinetic energy recovery and viscous dissipation of bouncing droplets undergoing head-on collision. The present study attempts to extend the problem from axisymmetric to three-dimensional (3D) and focus on the more challenging off-center droplet bouncing, which is more practical and general in real situations. In order to emphasize the influence of off-center collision (measured by the impact parameter) on droplet bouncing, the present study limits its scope to droplets of equal size to avoid possible complexity introduced by variable size ratio, although the size ratio effects have been demonstrated to be important to droplet collision in other studies [7,25,26,33,34].

The present study employs the 3D volume-of-fluid (VOF) method [35,36] to simulate off-center collisions of two droplets. To improve computational efficiency and avoid numerical coalescence caused by possible coarse mesh, we adopted the coupled level-set/VOF (CLSVOF) method with multiple maker functions [37-39] to track the liquid-gas interfaces. A major challenge of simulating droplet collision lies in the inability of the Navier-Stokes equations in capturing the rarified gas effects and the Van der Waals force [40] within the gas film between two colliding droplets, thus prohibiting the accurate prediction of droplet coalescence. Although Li [41] reported preliminary success in numerically predicting droplet coalescence by implementing Zhang and Law's solution [40] for the rarefied gas film, it is still computationally expensive to resolve the interface mesh to the scale of $O(10)$ nm, especially in 3D simulations. Fortunately, the current simulation of droplet bouncing does not involve treatments of interface rupture and is less numerically demanding than predicting droplet coalescence.

Vortex dynamics has been demonstrated to be of unique significance in understanding the physics of droplet collision [27,42-44], for example, vortex ring formation [43] in the coalescence of a droplet into a liquid pool [42] and vortex-induced internal mixing [27] upon the coalescence of two droplets. Following these studies, the present work seeks to numerically investigate the energy characteristics of two off-center colliding and bouncing droplets, and particularly focus on the vortex-dynamical interpretation of the off-center effects on viscous dissipation and kinetic energy recovery. The numerical methodology and validations are presented in Sec. II, followed by the results and discussions in Sec. III.

## II. NUMERICAL METHODOLOGY AND VALIDATIONS

### A. Numerical method

The continuity and incompressible Navier-Stokes equations,

$$\nabla \cdot \boldsymbol{u} = 0, \tag{1}$$

$$\rho(\partial \boldsymbol{u}/\partial t + \boldsymbol{u} \cdot \nabla \boldsymbol{u}) = -\nabla p + \nabla \cdot (2\mu \boldsymbol{D}) + \sigma \kappa \boldsymbol{n} \delta_s, \tag{2}$$

are solved by using the classic fractional-step projection method, where $\boldsymbol{u}$ is the velocity vector, $\rho$ the density, $p$ the pressure, $\mu$ the dynamic viscosity, and $\boldsymbol{D}$ the deformation tensor defined as $D_{ij} = (\partial_j u_i + \partial_i u_j)/2$. $\sigma \kappa \boldsymbol{n} \delta_s$ represents the surface tension term, where $\delta_s$ is a Dirac delta function, $\sigma$ the surface tension coefficient, $\kappa$ the local curvature, and $\boldsymbol{n}$ the unit normal vector local to the interface.

The present simulation uses the VOF method to track the free gas-liquid interface. To solve both the gas and liquid phases, the density and viscosity are constructed by the volume fraction $c$ as $\rho(c) = c\rho_l + (1-c)\rho_g$ and $\mu(c) = c\mu_l + (1-c)\mu_g$, in which the subscripts $l$ and $g$ denote the liquid and gas phases, respectively. The volume fraction $c$ satisfies the advection equation

$$\partial c/\partial t + \nabla \cdot (c\boldsymbol{u}) = 0 \qquad (3)$$

with $c = 1$ for liquid phase, $c = 0$ for gas phase, and $0 < c < 1$ for gas-liquid interface. The present VOF method had been implemented in the open source code, Gerris [35,36], featuring the 3D octree adaptive mesh refinement, the geometrical VOF interface reconstruction, the coupled level-set/VOF (CLSVOF) method with height-function curvature estimation, and the continuous surface tension formulation. Gerris has been demonstrated to be suitable for a wide range of multiphase problems [23,26,27,45].

As briefly discussed in the introduction, the simulation of droplet collision by using the conventional VOF approach is strongly influenced by the mesh resolution near the interface. Specifically, a coarse mesh would induce unphysical coalescence between different droplets that are supposed to bounce off each other with sufficiently refined mesh. Chen and Yang [45] developed a thickness-based adaptive mesh refinement method based on Gerris, and they could simulate droplet bouncing with a relatively smaller number of meshes because only the interface close to the interaction region was refined. However, the time step has to be decreased to accommodate numerical stability as the mesh is refined, thereby restricting the computational efficiency. Coyajee and Boersma [39] proposed a modified CLSVOF approach which applies different marker functions to describe separate interfaces and has been validated for droplet bouncing on a liquid film [46] and binary droplet collision [38]. The same idea of multiple marker functions was also implemented in Gerris by Hu *et al.* [37] to simulate binary droplet bouncing.

Following Coyajee and Boersma [39], we use two volume fractions in the present simulation, namely $c_1$ and $c_2$, to separately track the interface of each liquid droplet. As the two volume fractions independently describe the two droplets, the interfaces respectively belonging to the two droplets cannot contact and there is no intercorrelation between the two marker functions. Accordingly, the density and viscosity constructions for each droplet can be uniformly expressed as $\rho(c) = \rho(c_1) + \rho(c_2)$ and $\mu(c) = \mu(c_1) + \mu(c_2)$, respectively. It is noted that this method would always enforce droplet bouncing for any droplet collision process, and thus is applied in the present study for droplet collision cases that are known by experiment to result in bouncing.

### B. Numerical specifications

The 3D computational domain and numerical specifications are illustrated in Fig. 1. The domain is $6D$ in length and $4D$ in both width and height, and all boundaries are specified with the free outflow boundary conditions. Two droplets of diameter $D$ are specified to collide along the $x$-direction with a relative velocity, $U$, and zero velocities in the $y$- and $z$- directions. The $x$-velocity component for each droplet has the same magnitude of $U/2$ but opposite sign so that the linear momentum of the entire collision system remains zero. The *x-z* plane is established by the $x$-axis and the connection line denoted as $OO'$, in which $O$ and $O'$ are the mass centers of the colliding droplets. The midpoint of $OO'$ is located at the origin of the Cartesian coordinate system. It is noted that the *x-z* plane is always a plane of symmetry for the 3D colliding droplets. The deviation of the off-center collision from the head-on collision is qualified by $\chi$, which is defined as the projection of $OO'$ in the direction of the relative velocity.

Choosing $D$, $\rho_l$, and $\sigma$ as the basic units, we can nondimensionalize the eight relevant variables into five non-dimensional parameters, such as $We, Oh, B, \rho_g/\rho_l$, and $\mu_g/\mu_l$, where we have the Weber number, $We = \rho_l D U^2/\sigma$, the impact parameter, $B = \chi/D$, and the Ohnesorge number, $Oh = \mu_l/\sqrt{\rho_l \sigma D}$. In the present study concerning the collision between droplets in standard atmosphere environment, the gas-liquid density ratio and viscosity ratio are $\rho_g/\rho_l = 1.61 \times 10^{-3}$ and $\mu_g/\mu_l = 7.94 \times 10^{-3}$ (using tetradecane as an example), which has insignificant influence on droplet deformation and energy transfer according to previous studies [26,27]. The non-dimensional time is defined as $T = t/t_{osc}$, where $t$ is the physical time and $t_{osc}$ is the natural oscillation time defined as $t_{osc} = \sqrt{\rho_l D^3/\sigma}$.

To improve computational efficiency, the computational domain is divided into three physical zones, namely the gas, the liquid, and the droplet interface zone, and each zone has its own mesh refinement level denoted by $N$, which corresponds to a minimum mesh size of $O(2^{-N})$. Accordingly, $(N_1, N_2, N_3)$ is used to describe the refinement level in the three zones. A typical simulation run with the mesh refinement level (3, 5, 7) results in 514204 grid points, which is equivalent to about $2.0 \times 10^8$ grid points on a uniform mesh. It takes about 100 hours of real time to run the simulation up to $T = 2.0$ on an Intel Xeon(R) E5-2630 processor with 16 cores.

### C. Numerical validation and grid-dependence analysis

To validate the present numerical setup, the head-on droplet bouncing at two critical transition Weber numbers, corresponding to the so-called "soft" and "hard" collisions, and an off-center droplet bouncing are simulated and compared with the experimental results from Pan *et al.* [5] and Qian and Law [6], respectively. Fig. 2 shows the comparison between experimental images (left) and simulation results (right). The simulation results of droplet deformation closest to the experimental images are presented, whereas the time discrepancies between simulation and experiment could serve as an indicator for the simulation performance.

Fig. 2(a) shows the head-on droplet bouncing at $We = 2.3$ and $Oh = 2.80 \times 10^{-2}$ ($t_{osc} = 1.06 ms$) [5]. The experimental and simulation times are nearly identical in early collision stages and begin to display slight discrepancies as time evolves in later stages. The time errors are generally less than 3% except it is about 5% at T = 1.08. Similarly, Fig. 2(b) shows the head-on droplet bouncing at a larger Weber number with $We = 9.3$ and $Oh = 2.78 \times 10^{-2}$ ($t_{osc} = 1.03 ms$) [5]. Again, the time errors are less than 2%.

Fig. 2(c) shows the off-center droplet bouncing adapted from Fig. 4(r) in Qian and Law [6], who reported this case with $We = 14.5, Oh = 2.80 \times 10^{-2}$ and $B = 0.34$. After careful analysis of their experimental images, we found that the droplets actually undergo nearly grazing collision with $B = 0.9$ instead of 0.34. It is likely that the original authors made a typo by accidentally swapping the annotations between Fig. 4(r) and Fig. 4(q) of [6], so that using $We = 48.8, Oh = 2.80 \times 10^{-2}$ and $B = 0.82$ enabled quantitative reproduction of all the collision images. The time errors are about 10%, which is probably attributed to the experimental uncertainties in the measurement for time and size [6]. Our findings can also be justified by the simulation of Chen and Yang [45], who used $We = 48.8, Oh = 2.93 \times 10^{-2}$ and $B = 0.9$, as shown in Fig. 2(d). It is seen that the simulations (c) and (d) predict similar droplet deformation although the impact parameters are slightly different.

The grid-independence analysis was performed for the validation case (a) of Fig. 2. The kinetic energy (KE) and surface energy (SE) are normalized by the initial total energy and compared for totally six different mesh refinement level sets ($N_1, N_2, N_3$), as shown in Fig. 3. By fixing the mesh refinement level of the gas and the liquid zones at $N_1 = 2$ and $N_2 = 4$, respectively, we can analyze the grid dependence of the droplet interface zone, and the results show convergence up to T = 1.2 as increasing the mesh refinement level from $N_3 = 7$ to $N_3 = 8$. Similarly, the result comparison between (2, 4, 7) and (3, 5, 7) and further comparison between (2, 4, 8) and (4, 6, 8) also imply convergence of simulation with different mesh refinement levels of the gas and the liquid zones. For a balance between computational cost and accuracy, the intermediate mesh refinement level of (3, 5, 7) has been used in the validation cases and all other simulations in the following sections.

## III. RESULTS AND DISCUSSIONS

### A. Phenomenological description

The representative case of $We = 9.3$ and $Oh = 2.80 \times 10^{-2}$ at $B = 0.0$ and $B = 0.3$ has been used to phenomenologically describe the differences between the head-on and the off-center droplet bouncing, as shown in Fig. 4(a) and 4(b), respectively. The projection of droplet deformation on symmetry (*x-z*) plane is shown in the first row of Fig. 4(a) and 4(b), in which the mass of the colliding droplets centers, $O$ and $O'$, are indicated as black solid points. To analyze the flow field within the droplets, the pressure contours and streamlines on the symmetry (*x-z*) plane and on the $YOO'$ plane where the y-axis and the line $OO'$ lie, are shown in in Fig. 4(b) for the case of $B = 0.3$.

The droplet bouncing process is similar between head-on and off-center collisions in some aspects. First, as shown by the streamlines, the colliding droplets deform from their initial spherical shape to the maximum "dumbbell-like" shape (about T = 0.38), which results in the locally enhanced capillary pressure around large curvatures. Then, the deformed droplets are driven by the capillary pressure difference to bounce back, meanwhile the surface energy converts back to the kinetic energy. The bouncing droplets generally experience several oscillation periods before recovering their spherical shape.

It is noted that the pressure distributions and streamlines on $YOO'$ plane for head-on and off-center collisions are qualitatively similar except the pressure amplitude for the off-center collision is slightly smaller. That is because the $YOO'$ plane is perpendicular to the symmetry *x-z* plane, and thereby the pressure distributions and streamlines on $YOO'$ plane should be symmetric for each droplet. The distinctive differences can be observed on the *x-z* plane for the off-center collision that the mass deviation in *z*-direction results in the asymmetric pressure distribution for each droplet, where the unbalanced capillary pressure tends to promote droplets oscillation and delay droplets recovery to their spherical shape, as the contours at T = 0.53 and T = 0.70 shown in Fig. 4. In addition, the vortical flow is clearly seen around the droplet interaction region only for the off-center collision on *x-z* plane but not on $YOO'$ plane and both planes for the head-on collision.

**B. Mass center trajectory and interface hysteresis**

To further understand the off-center droplet bouncing, the above comparison between $B = 0.0$ and $B = 0.3$ has been extended to include two more impact parameters, namely $B = 0.6$ and 0.9. The whole collision process can be quantitatively divided into four stages based on the mass center distance $D_{OO'}$, as the imbedded picture shown in Fig. 5(a). As the colliding droplets approaching, the time instant corresponding to $D_{OO'} = D$ (for the first time) is defined as T = 0, and then the end of Stage I, II, and III are defined as $D_{OO'}$ reaching a minimum, recovering back to $D_{OO'} = D$, and bouncing away to $D_{OO'} = 3D$, respectively. It is seen that the time durations of Stage I, II, and III tend to decrease with increasing $B$. However, the time duration of $B = 0.3$ is slightly longer than that of $B = 0.0$ in Stage II, and this difference of time duration has been further enlarged substantially in Stage III, indicating the droplet entanglement time is prolonged for moderately off-center collisions and showing a non-monotonic variation with increasing $B$.

The trajectories of droplets mass centers are always lie on the *x-z* symmetry plane, as shown in Fig. 5(b). The initial mass center positions of colliding droplets for different impact parameters are located on a circle, whose diameter equals to the mass center distance $D_{OO'} = D$. The mass center trajectory is along the *x*-direction for the head-on collision but deflected for the off-center collisions. This deflection referring to the change of velocity vector is uniquely determined by $B$ (for fixed $We$ and $Oh$) based on the conservation laws of momentum, angular momentum, and energy.

As an interesting phenomenon of bouncing droplets, the interface hysteresis [5,32] describes the colliding interfaces of two bouncing droplets tend to approach each other for a short period of time when the mass centers of droplets have already started to depart from each other. The previous study [32] on bouncing droplets after head-on collisions found that smaller droplet deformation by decreasing the size ratio, or by decreasing $We$, or by increasing $Oh$, favors the interface hysteresis. As shown in Fig. 5(c), the interface hysteresis can also be observed for bouncing droplets after off-center collisions. The defined interface gap distance $\delta_{gap}$, which describes the minimum distance of two interfaces in the direction of the mass center connection line $OO'$, is illustrated in the imbedded picture in Fig. 5(a). Furthermore, it is seen a most prominent drop of $\delta_{gap}$ for $B = 0.3$. This is probably attributed to the decreased bouncing velocities of droplets after moderately off-center collisions. A slow movement of bouncing interfaces favors the interface hysteresis, which allows a sufficient time for the pressure relaxation within the compressed gas film so that the higher pressure on the liquid side pushes the concave (toward the droplets) interface to be convex (toward the gas film) [5,32].

### C. Energy budget analysis

Fig. 6(a) shows the time evolution of the kinetic energy (KE) and the surface energy (SE) of the liquid droplets, which are normalized by their initial total energy (KE+SE). During Stage I involving the droplet deformation, the total decrement of KE and the total increment of SE decrease monotonically with increasing $B$ from 0.0 to 0.9. This is because increasing $B$ tends to decrease the effective impact velocity, $U\sqrt{1-B^2}/2$, of colliding droplets along the mass center connection line and thereby results in smaller droplet deformation with a smaller KE loss. Furthermore, it is noted that the occurrence of minimum mass center distance at the end of Stage I, as shown by the open circles, is not synchronous with the peak value of SE and the local minimum of KE. This means the droplets squeezing process has not been accomplished when the droplet mass centers start to bounce away. This can be also treated as an interface hysteresis phenomenon.

In Stage II, apart from the short period of droplet squeezing process in the beginning, the droplets start to bounce back with a decrease of SE and an increase of KE. In Stage III, it is noted that there are several apparent oscillations of SE and KE especially for $B = 0.3$ and $B = 0.6$, and these oscillations are gradually attenuated by the viscous dissipation. An interesting observation during the Stage III is that the curve corresponding to the KE for $B = 0.3$ is always below those for other $B$s. This can be explained by that more KE has been transferred into the SE owing to the enhanced droplet oscillation.

To further explain this observation, the total viscous dissipation energy (TVDE) normalized by the initial total energy is shown in Fig. 6(b). The TVDE is defined by

$$\text{TVDE(T)} = \int_0^T \left( \int_{V_l} \phi \, dV \right) dT' \quad (4)$$

where $V_l$ is the volume of liquid droplets and $\phi$ is the local viscous dissipation rate (VDR) given by [47]

$$\phi = \mu \left[ 2\left(\frac{\partial u}{\partial x}\right)^2 + 2\left(\frac{\partial v}{\partial y}\right)^2 + 2\left(\frac{\partial w}{\partial z}\right)^2 + \left(\frac{\partial u}{\partial y} + \frac{\partial v}{\partial x}\right)^2 + \left(\frac{\partial v}{\partial z} + \frac{\partial w}{\partial y}\right)^2 + \left(\frac{\partial w}{\partial x} + \frac{\partial u}{\partial z}\right)^2 \right]. \quad (5)$$

It is noted that the total energy (TE) of the liquid droplets is close to unity during the entire collision process, indicating the energy budget of the gas flow is insignificant compared with that of the droplets.

As shown in Fig. 6(b), the TVDE of $B = 0.3$ at the end of Stage III is the largest among all the $B$s, indicating that the non-monotonic kinetic energy recovery is caused by the non-monotonic TVDE. To further understand the viscous dissipation effects on bouncing droplets in different stages, the temporal total viscous dissipation rate will be discussed in Section D.

### D. Enhanced viscous dissipation for moderately off-center collisions

The total viscous dissipation rate (TVDR), defined as a volume integral $\int_{V_l} \phi \, dV$, and the contours of local VDR on the symmetry (*x-z*) plane are illustrated in Fig. 7(a) and 7(b), respectively, to explain the enhanced viscous dissipation at $B = 0.3$. It is seen a prominent peak value of TVDR in Stage I and Stage II, respectively, but several local maxima in Stage III. Three representative time instants, $T_1$, $T_2$ and $T_3$, correspond to three local maximums of TVDR.

First, for each $B$, the TVDR during Stage I is larger than those during other stages and makes a major contribution on the entire process of viscous dissipation. As shown by a representative time instant, $T_1$, in Fig. 7(b), the local VDR is mainly distributed around the droplets interaction region with a rapid droplet expansion, showing a boundary-layer-like internal flow that was also observed in the previous theoretical analysis [40] and numerical simulation [32]. Furthermore, for the comparison between different $B$s during Stage I, it is seen that the TVDR decreases with increasing $B$. Due to the droplet deformation in Stage I is inertia-dominant, following Jiang *et al.* [4] and Tang *et al.* [7], the observation can be understood by an approximate estimation of the TVDE as $\alpha E_k(B) = \alpha(1 - B^2)E_{k0}$ where $\alpha$ is a viscous dissipation coefficient and $E_k(B)$ is the effective impact KE of colliding droplets along the mass center connection line.

Second, during Stage II, the droplets bounce back under the capillary pressure difference with the SE transferring into the KE. As shown by a representative time instant, $T_2$, in Fig. 7(b), the VDR in the droplet interior being away from the interaction region decreases from maximum at $B = 0.0$ to nearly vanishing at $B = 0.9$. This is because, based on a scaling relation $\Delta\text{SE} \sim \Delta\text{KE} \sim \frac{1}{2}mu^2$ where $\Delta\text{SE}$ and $\Delta\text{KE}$ are the change of SE and KE during each stage, respectively, the TVDR can be estimated as $2\mu_l(\Delta\text{SE})V_l/mD^2$, where the characteristic velocity is $u \sim \sqrt{2\Delta\text{SE}/m}$ and the characteristic length is $D$. It indicates that the larger droplet deformation in Stage I directly accounts for the larger $\Delta\text{SE}$ and thereby enhanced TVDR in Stage II. Furthermore, it is interesting to find that the VDR in the vicinity of the droplet interacting region is observed for off-center collisions but not for the head-on collision. This part of VDR is owing to the stretching effect of off-center collisions, and its amplitude seemingly increases as increasing $B$. Consequently, the TVDR of $B = 0.3$ is enhanced and becomes the largest among these $B$s by the competition mechanism between these two parts of VDR concentration. The two parts of VDR can be clearly distinguished in Fig. 7(b) by blanking the contours with a low threshold value of 0.5.

Third, as the local maxima of the TVDR shown in Fig. 7(a), the TVDR during Stage III is further enhanced for $B = 0.3$, which is attributed to its stronger droplet oscillation illustrated by the change of SE in Fig. 6(a). In addition, the enhanced droplet oscillation for moderately off-center collisions is probably induced by the unbalanced capillary pressure distributions, as shown in Fig. 4(b).

### E. Non-monotonic Kinetic energy recovery

To compare droplet bouncing at $We = 2.3$ and $We = 5.8$, the droplet KE and TVDE are shown in Fig. 8(a) and 8(b), respectively. The normalized KE and TVDE and their oscillation amplitudes are all reduced as decreasing $We$ from 9.3 to 2.3. This is because the decreased KE evokes a smaller change of SE so that the oscillations of SE and KE damp more rapidly. However, it is noted that the non-monotonic variation of TVDE and thereby kinetic energy recovery with increasing $B$ can be observed for all these $We$s.

In Lagrangian simulation of sprays [48-52], the kinetic energy dissipation factor, defined as $f_E = 1 - \text{KE}'/\text{KE}$, where KE and KE′ are the kinetic energy of droplets before and after collisions, respectively, is required to determine the post-collision velocities of droplets based on the conservation laws of momentum, angular momentum, and energy. In the present study, the KE before collision is determined by the initial state, while for consistency among all $B$s the KE′ after collision is evaluated by the time instant at the end of Stage III. Accordingly, the kinetic energy recovery factor can be defined as $1 - f_E$.

Fig. 9 shows $f_E$ with varying $B$s at different $We$s and $Oh$s. Fig. 9(a) shows $f_E$ with varying $B$s at different $We$s. Among the small and intermediate $B$s in the range of 0.0 ~ 0.7, $f_E$ increases as increasing $We$ and the deviations of $f_E$ between different $We$s are enlarged at intermediate $B$.

This is because the viscous dissipation induced by the droplet deformation is enhanced as increasing $We$. However, it is interesting to find that, among the large $B$s in the range of 0.7 ~ 1.0 for near grazing collisions, the trend has been reversed so that $f_E$ decreases as increasing $We$. This can be understood as that, although increasing $We$ can generally enhance the droplet deformation and internal flow, it reduces the time of droplet interaction for viscous dissipation and therefore leads to the decreased $f_E$ at large $B$s. Furthermore, the critical impact parameter $B_{cr}$, which is defined as the $B$ with maximum $f_E$ (for fixed $We$ and $Oh$), decreases slightly as increasing $We$. Fig. 9(b) shows $f_E$ with varying $B$s at different $Oh$s. It is seen that $f_E$ increases as increasing $Oh$ for all $B$s, which is easily understood as the enhanced viscosity. The largest difference of $f_E$ between different $Oh$s occurs at $B = 0.0$. The $B_{cr}$ is about 0.3 and seemingly unaffected by $Oh$.

Overall, the non-monotonic kinetic energy recovery can be treated as a general phenomenon for different $We$s and $Oh$s. To serve for Lagrangian spray simulations, Zhang and Zhang [32] had conducted a comprehensive parametric study on the bouncing droplets after head-on collisions and proposed a fitting formula correlating $f_{E0}(We, Oh) = f_E(We, Oh, B = 0)$ with $We$ and $Oh$. Following their conclusions, for the present study by considering additional impact parameter effects, an approximate fitting formula of

$$f_E/f_{E0} = a(B - B_{cr})^2 + b \tag{6}$$

can be proposed for the concerned $We = 2.3$~$9.3$ and $Oh = 1.4 \times 10^{-2}$~$5.6 \times 10^{-2}$, where $B_{cr} = 0.5 - 0.2(We/10)$, $b = 0.83 + 0.3(We/10) + 0.9(Oh/10^{-2})^{-1}$ and $a = (1 - b)/B_{cr}^2$. It reflects the most important finding of the non-monotonic kinetic energy recovery with varying $B$ for various $We$s and $Oh$s. More details about the analysis of this approximate fitting formula are given in Supplemental Material [53].

## F. Vortex lines and Helicity

As shown in Fig. 7(b), the VDR in the vicinity of droplet interaction region can be observed for off-center collisions, and it is accompanied by the occurrence of vortical flows, which are not observed in head-on collisions. Thus, the present section aims to study the different vortical flows between head-on and off-center droplet collisions.

Fig. 10(a) shows a qualitative comparison of vortex lines between $B = 0.0$ and $B = 0.3$ at four different time instants. It is seen that the vortex lines for the head-on collision are a series of concentric circles, which are centered along the line connecting the mass centers of the droplets, indicating that the flow is purely axisymmetric and the vortices are in the ring shape. These ring-shaped vortices should be attributed to the shear flow formed by compression in the axial direction and expansion in the radial direction. For the off-center collision, the "ring-shaped" vortex lines can still be observed in the droplet interior being away from the interaction region between the two droplets. An interesting finding is the "line-shaped" vortex lines in the vicinity of the droplet interaction region, as the zoom-in picture shown in Fig. 10(b), which corresponds to a narrow viscous shear layer generated by the stretching effect between the two droplets. These vortex lines imply that the direction of the shear layer is parallel to both the *x-z* plane and the droplets interacting surface. We can further observe that both ends of a "line-shaped" vortex line stem from the droplet surface. This can be explained by that, in a finite vorticity field, a vortex filament must either form a closed vortex ring or terminate on the fluid boundaries where the flux of vorticity is not zero [54].

The vortical structures and flow characteristics of off-center collisions can be further understood by considering the "non-orthogonality" of the velocity and vorticity vectors, which is conventionally measured by helicity, which is defined as

$$H = \int \vec{u} \cdot \vec{\omega} dV \qquad (7)$$

where $\vec{u} = (u, v, w)$ is the velocity vector and $\vec{\omega}$ is the vorticity vector given by

$$\vec{\omega} = \nabla \times \vec{u} = \left(\frac{\partial w}{\partial y} - \frac{\partial v}{\partial z}, \frac{\partial u}{\partial z} - \frac{\partial w}{\partial x}, \frac{\partial v}{\partial x} - \frac{\partial u}{\partial y}\right). \qquad (8)$$

Evidently, $\vec{u} \cdot \vec{\omega}$ remains zero for head-on collisions because the axisymmetric flow has only one vorticity component normal to the velocity plane. However, the defined helcity is zero if the integration is over the entire droplets. Because the *x-z* plane is a symmetry plane, the velocity components and the velocity derivatives are reversed on the both sides of the *x-z* plane, expressed as $v_+ = -(v_-)$ and $(\partial/\partial y)_+ = -(\partial/\partial y)_-$ where the positive and negative symbol denote two sides of *x-z* plane, respectively. Thus, we have

$$(\vec{u}_- \cdot \vec{\omega}_-) = (u, -v, w) \cdot \left(-\frac{\partial w}{\partial y} + \frac{\partial v}{\partial z}, \frac{\partial u}{\partial z} - \frac{\partial w}{\partial x}, -\frac{\partial v}{\partial x} + \frac{\partial u}{\partial y}\right) = -(\vec{u}_+ \cdot \vec{\omega}_+) \qquad (9)$$

and $H = H_+ + H_- = 0$ for the entire droplets.

Based on the above consideration, we defined the "helicity" of liquid droplets in a half space as

$$H_{1/2} = |H_+| = \left|\int_{V_{l_+}} (\vec{u}_+ \cdot \vec{\omega}_+) dV\right|. \qquad (10)$$

As shown in Fig. 11(a), $H_{1/2}$ are enhanced for intermediate $B$ values of 0.3 and 0.6, showing a non-monotonic variation of $H_{1/2}$ as increasing $B$. This can be understood as that, at intermediate $B$s, both the "ring-shaped" vortices and the "line-shaped" shear layer present strong influence on their surrounding flows, and their interations are reflected by the strong linkage or knotedness of the vortical strucutres, which translates into strong helicity in the current context. For the cases of small or large $B$, the "ring-shaped" vortices and the "line-shaped" shear layer are reduced or weakened, thus the entanglement between different types of vortices is suppressed, resulting in the marginal or reduced helicity.

To further demonstrate the source of helicity, the vortex lines, streamlines, and local distribution of the density of helicity, $\vec{u} \cdot \vec{\omega}$, at several cross sections parallel to the droplet interacting surface are plotted in Fig. 11(b), showing the transition of vortex lines from "line-shaped" to "ring-shaped". The results show that the nonzero $\vec{u} \cdot \vec{\omega}$ is mainly distributed near the droplet interacting surface or other interfaces, where shear layer structure (characterized by the "line-shaped" vortex lines) encounters radial flow induced by the "ring-shaped" vortices.

### G. Correlation between TVDR and total enstrophy

The effect of vorticity on viscous dissipation will be discussed in this section by analyzing the correlation between the enstrophy dissipation rate, following Tran and Dritschel's definition in [55], $\epsilon = \mu \omega^2$, and the viscous dissipation rate $\phi$. The volume integrals $\int_{V'} \phi \, dV$ and $\int_{V'} \epsilon \, dV$, which are hereafter denoted as $\Phi$ and $E$, respectively, are known to be equal for any single-phase flow in a closed domain $V'$ confined by stationary boundaries $S'$ [56]. This apparently does not guarantee the local equivalence between $\epsilon$ and $\phi$ everywhere inside $V'$. Furthermore, we recognized that the equivalence between $\Phi$ and $E$ does not hold for a domain containing free two-phase interfaces. Specifically, for the present problem involving liquid-gas interfaces $S_l$, we have derived the correlation between $\Phi$ and $E$ in Appendix A to be

$$\Phi - E = \int_{V'} (\phi - \epsilon) \, dV = \int_{S_l} \big( ([\![p]\!]_S - \sigma \kappa) u_n + \mu_l \omega_l u_t \big) dA, \tag{11}$$

where $[\![p]\!]_S$ denotes the pressure jump across the interface $S_l$, $u_n$ and $u_t$ are the normal and tangential velocity components, respectively, and $\omega_l$ is the vorticity in the liquid phase. It should be noted that the result of Eq. (11) has been simplified based on the approximation of negligible gas viscosity, as discussed in Section C.

Eq. (11) can be understood that the difference between the total viscous dissipation and the total enstrophy for two-phase flow is attributed to two sources at the liquid-gas interface. One is the work done by the normal component of the viscous stress, which equals $[\![p]\!]_S - \sigma\kappa$ as explained in Appendix A, and the other is the work done by a virtual shear stress, $\mu_l \omega_l$. In this study, the outer air phase is insignificant and has negligible effect on momentum or energy transfer owing to the large liquid-to-gas ratios of density and viscosity. Thus, the free outflow boundaries in simulations may be considered as approximately stationary boundaries without losing much accuracy in the liquid phase.

Fig. 12(a) presents the comparison of numerical result of $\Phi - E$ for four different $B$s experiencing several periods of oscillation during droplet collision. Snapshots of contours of local $\phi$ and $\epsilon$ in the $x$-$z$ plane are plotted at the instants corresponding to local maxima and minima of the main curve for the case of $B = 0.3$. Fig. 12(a) verifies that $\Phi - E$ is indeed non-zero for the present two-phase flow problem as predicted by Eq. (11). It can be further observed that $\Phi - E$ oscillates and the amplitude decreases gradually after several periods of droplet oscillation, in a manner similar to the evolution of TVDR shown in Fig. 7, and with identical time instants for each local extrema. It is therefore implied that, also be consistent with the interpretation from Eq. (11), $\Phi - E$ is closely related to the interfacial movement, which is synchronized with the oscillation of the entire droplet.

To further verify the connections between $\Phi - E$ and droplets interfacial movement, we plotted the change rates of the normalized KE and SE, in Fig. 12(b). It shows that the KE and the SE always change synchronously during the droplets collision process, and their difference is the viscous dissipation rate. For the head-on collision, the time instants with $\Phi - E = 0$ in Fig. 12(a) and those with $d(\text{SE})/dt = -d(\text{KE})/dt = 0$ corresponding to the maximum droplets deformation in Fig. 12(b) are approximately matched, indicating that $\Phi - E = 0$ occurs at maximum droplet deformation, and thereby nearly stationary interfacial movement. However, the mismatch between the time instants with $d(\text{SE})/dt = -d(\text{KE})/dt = 0$ and $\Phi - E = 0$ for off-center collisions is probably attributed to the fact that maximum droplet deformation does not guarantee stationary interfacial movement everywhere for any oscillating droplets, hence resulting in a hysteresis between the zero points of Fig. 12(a) and (b).

## IV. CONCLUDING REMARKS

Two droplets undergoing off-center collisions were simulated in the present study by using a modified CLSVOF approach and validated against experiments of both head-on and off-center collisions. The implementation of multiple marker functions helped to avoid the unphysical numerical coalescence which could occur under relatively coarse mesh of the interface. The analysis of mass center trajectory demonstrates that the interface hysteresis is enhanced and the droplet entanglement time is prolonged for moderately off-center collisions. A non-monotonic kinetic energy recovery with varying impact parameter was observed, which is attributed to the enhanced viscous dissipation of moderately off-center collisions. Specifically, the VDR in the droplet interior away from the interaction region decreases with increasing $B$, whereas the VDR in the vicinity of the droplet interacting region increases owing to the droplet stretching. The competition mechanism between these two parts of VDR accounts for the enhanced viscous dissipation at intermediate $B$ in the early stages of droplet collision; while the enhanced droplet oscillation owing to the unbalanced capillary pressure distributions results in the further increase of viscous dissipation at intermediate $B$ in the late periods of droplet oscillation. The non-monotonic kinetic energy recovery has been quantitatively verified to be a general phenomenon for various $We$ and $Oh$, and an approximate fitting formula, $f_E/f_{E0} = a(B - B_{cr})^2 + b$, has been proposed for $We = 2.3 \sim 9.3$ and $Oh = 1.4 \times 10^{-2} \sim 5.6 \times 10^{-2}$.

The helicity analysis shows that the helicity is zero for the head-on collision because of the axisymmetric flow with only "ring-shaped" vortices, but the helicity is nonzero for off-center collisions owing to the presence of vortex interactions between the "ring-shaped" vortices and the "line-shaped" shear layers. The helicity also non-monotonically varies with increasing $B$. Furthermore, in the present two-phase flow system, the total enstrophy was found to be different from the total viscous dissipation, which contradicts the equivalence between the two terms for a single-phase flow. This deficit is analytically attributed to two terms originated from the phase interface: the work done by the viscous stress along the normal direction and the work done by a virtual shear stress along the tangential direction. Both terms are caused by the unbalanced flow and vorticity across the interface and are found to be closely related to the interfacial movement.

## ACKNOWLEDGMENTS

The work was supported partly by the Hong Kong RGC/GRF (PolyU 152217/14E and PolyU 152651/16E) and partly by the Hong Kong Polytechnic University (G-YBXN). We are grateful to State Key Laboratory of Engines of Tianjin University for an "Open Fund" (No. K2018-12) and to Dr. Zhenyu Zhang for his insightful advices for result discussion.

## APPENDIX A

Following the derivations in [56], we found that for a multiphase flow containing free interface the relationship between the viscous dissipation rate $\phi$ and the enstrophy dissipation rate $\epsilon$ could be expressed as

$$\phi - \epsilon = \nabla \cdot (\boldsymbol{u} \cdot \boldsymbol{\tau}) - 2\boldsymbol{u} \cdot \boldsymbol{D} \cdot \nabla\mu - (\boldsymbol{u} \times \nabla\mu) \cdot \boldsymbol{\omega} - \nabla \cdot (\mu \boldsymbol{u} \times \boldsymbol{\omega}), \tag{A1}$$

where the viscous stress tensor $\boldsymbol{\tau}$ is related to the strain rate tensor $\boldsymbol{D}$ as $\boldsymbol{\tau} = 2\mu\boldsymbol{D}$. We next calculated the volume integral of Eq. (A1) for the present two-phase flow system shown in Fig. A1, where $V'$ is the volume of the entire domain including both gas and liquid phases, and $V_l$ denotes the volume occupied by the liquid droplets enclosed by surface $S_l$. $V'$ is assumed to be large enough such that the flow becomes stagnant at the domain boundary $S'$.

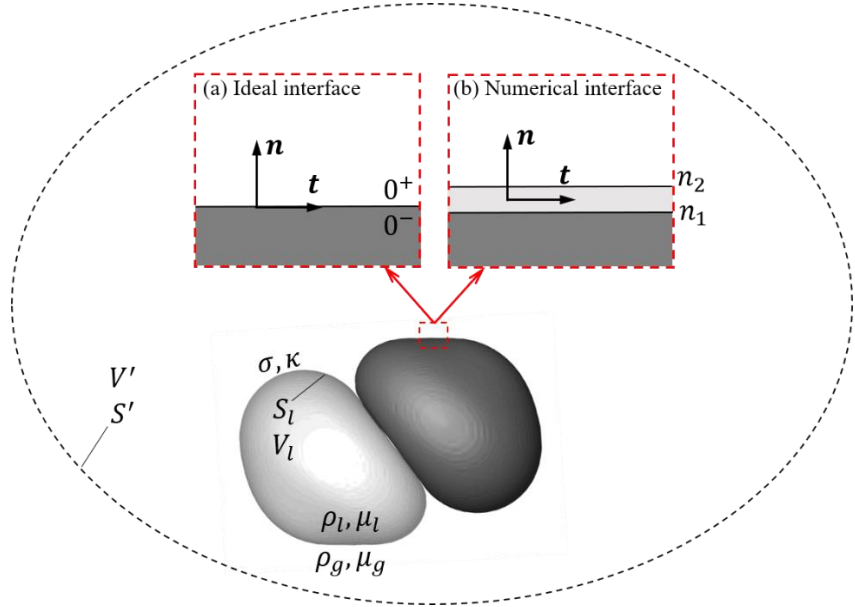

FIG. A1. The schematic of a two-phase flow system with different models for the liquid-gas interface.

First, we considered the liquid-gas interface $S_l$ as an ideal mathematical interface with discontinuous properties across it, as illustrated in Fig. A1(a). Then, the integrals associated with $2\boldsymbol{u}\cdot\boldsymbol{D}\cdot\nabla\mu$ and $(\boldsymbol{u}\times\nabla\mu)\cdot\boldsymbol{\omega}$ vanish because $\nabla\mu=0$ is satisfied in both liquid- and gas phases constituting the entire integral domain $V'$. To obtain the volume integrals of $\nabla\cdot(\boldsymbol{u}\cdot\boldsymbol{\tau})$ and $\nabla\cdot(\mu\boldsymbol{u}\times\boldsymbol{\omega})$, we noted that the Gauss theorem in the present case takes the form of

$$\int_{V'}\nabla\cdot\boldsymbol{F}\,\mathrm{d}V = \oint_{S'}\boldsymbol{F}\cdot\boldsymbol{n}\,\mathrm{d}A - \oint_{S_l}[\![\boldsymbol{F}\cdot\boldsymbol{n}]\!]_S\,\mathrm{d}A, \tag{A2}$$

where the jump of a function $g$ is defined as $[\![g]\!]_S = g(n=0^+) - g(n=0^-)$, with $n=0$ denoting the location of the free interface $S_l$, as shown in Fig. A1(a).

Applying Eq. (A2), the volume integral of $\nabla\cdot(\boldsymbol{u}\cdot\overleftrightarrow{\boldsymbol{\tau}})$ can be derived as

$$\int_{V'}\nabla\cdot(\boldsymbol{u}\cdot\boldsymbol{\tau})\,\mathrm{d}V = -\oint_{S_l}[\![\boldsymbol{u}\cdot\boldsymbol{\tau}\cdot\boldsymbol{n}]\!]_S\,\mathrm{d}A = \oint_{S_l}([\![p]\!]_S - \sigma\kappa)u_n\,\mathrm{d}A, \tag{A3}$$

where the surface integral on $S'$ is eliminated because of the stagnation boundary condition. The second equation of (A3) is derived by taking advantage of the stress boundary conditions across the free interface $S_l$ that $[\![\boldsymbol{t}\cdot\boldsymbol{\tau}\cdot\boldsymbol{n}]\!]_S = \boldsymbol{0}$ and $[\![\boldsymbol{n}\cdot\boldsymbol{\tau}\cdot\boldsymbol{n}]\!]_S = -[\![p]\!]_S + \sigma\kappa$ [57], where $\boldsymbol{n}$ and $\boldsymbol{t}$ respectively denotes the unit vectors in the normal and tangential directions of $S_l$. So $u_n = \boldsymbol{u}\cdot\boldsymbol{n}$ is the normal velocity component of $S_l$.

Similarly, the integral of $\nabla\cdot(\mu\boldsymbol{u}\times\boldsymbol{\omega})$ can be derived as

$$\int_{V'}\nabla\cdot(\mu\boldsymbol{u}\times\boldsymbol{\omega})\,\mathrm{d}V = -\oint_{S_l}[\![\mu(\boldsymbol{u}\times\boldsymbol{\omega})\cdot\boldsymbol{n}]\!]_S\,\mathrm{d}A = \oint_{S_l}[\![\mu\omega]\!]_S u_t\,\mathrm{d}A, \tag{A4}$$

where $\boldsymbol{\omega} = \omega(\boldsymbol{t} \times \boldsymbol{n})$, and $u_t = \boldsymbol{u} \cdot \boldsymbol{t}$ is the velocity component in the tangential direction of interface $S_l$. It is noted that the derivation of Eq. (A4) requires the no-slip boundary condition at the liquid-gas interface so that $u_t$ is continuous across $S_l$.

We combined Eqs. (A1), (A3), and (A4) to obtain

$$\int_{V'} (\phi - \epsilon) \, dV = \oint_{S_l} \left( (\llbracket p \rrbracket_S - \sigma \kappa) u_n + \mu_l \omega_l u_t \right) dA, \tag{A5}$$

where $\omega_l$ is the vorticity of the liquid phase generated from the free interface [27,58]. The $\mu \omega u_t$ term pertaining to the gas phase is dropped out because of the negligible gas viscosity compared with fluid viscosity.

We noted that the interface between different phases processed by numerical simulation is often modeled as a finite-thickness surface with continuously varying properties in the normal direction, as illustrated in Fig. A1(b). Thus, the interface location in the normal direction is extended from $n = 0$ to $[n_1, n_2]$ where $n_1 < 0$ to $n_2 > 0$. In this case, the Gauss theorem has its original form so that the volume integrations of $\nabla \cdot (\boldsymbol{u} \cdot \boldsymbol{\tau})$ and $\nabla \cdot (\mu \boldsymbol{u} \times \boldsymbol{\omega})$ vanish again [56]. However, since $\nabla \mu$ now becomes finite throughout the finite-thickness interface, the integrations of $2\boldsymbol{u} \cdot \boldsymbol{D} \cdot \nabla \mu$ and $(\boldsymbol{u} \times \nabla \mu) \cdot \boldsymbol{\omega}$ can be respectively derived as

$$\int_{V'} 2\boldsymbol{u} \cdot \boldsymbol{D} \cdot \nabla \mu \, dV = \oint_{S_l} \int_{n_1}^{n_2} 2\boldsymbol{u} \cdot \boldsymbol{D} \cdot \boldsymbol{n} \frac{\partial \mu}{\partial n} dn dA = \oint_{S_l} \llbracket \boldsymbol{u} \cdot \boldsymbol{\tau} \cdot \boldsymbol{n} \rrbracket_S \, dA \tag{A6}$$

and

$$\int_{V'} (\boldsymbol{u} \times \nabla \mu) \cdot \boldsymbol{\omega} \, dV = \oint_{S_l} \int_{n_1}^{n_2} (\boldsymbol{u} \times \boldsymbol{n}) \cdot \boldsymbol{\omega} \frac{\partial \mu}{\partial n} dn dA = - \oint_{S_l} \llbracket \mu (\boldsymbol{u} \times \boldsymbol{\omega}) \cdot \boldsymbol{n} \rrbracket_S \, dA, \tag{A7}$$

where the jump of $g$ here is defined as $[\![g]\!]_S = g(n = n_2) - g(n = n_1)$. Combining Eqs. (A1), (A6), and (A7), we again attained Eq. (A5).

---

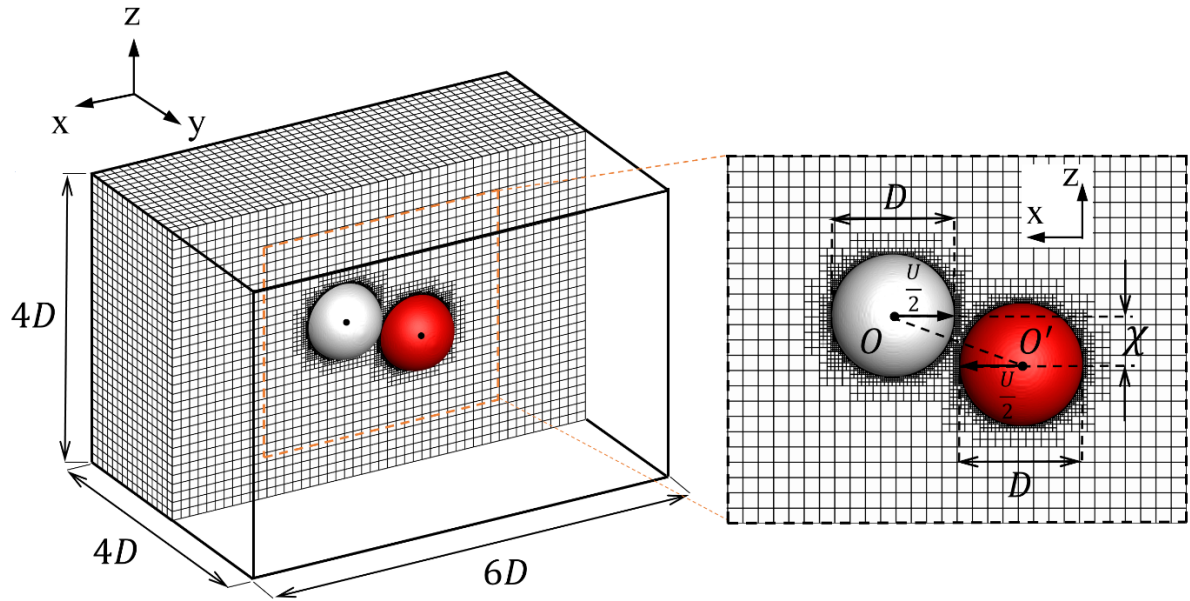

FIG. 1. Computational domain and setup for off-center droplet collision.

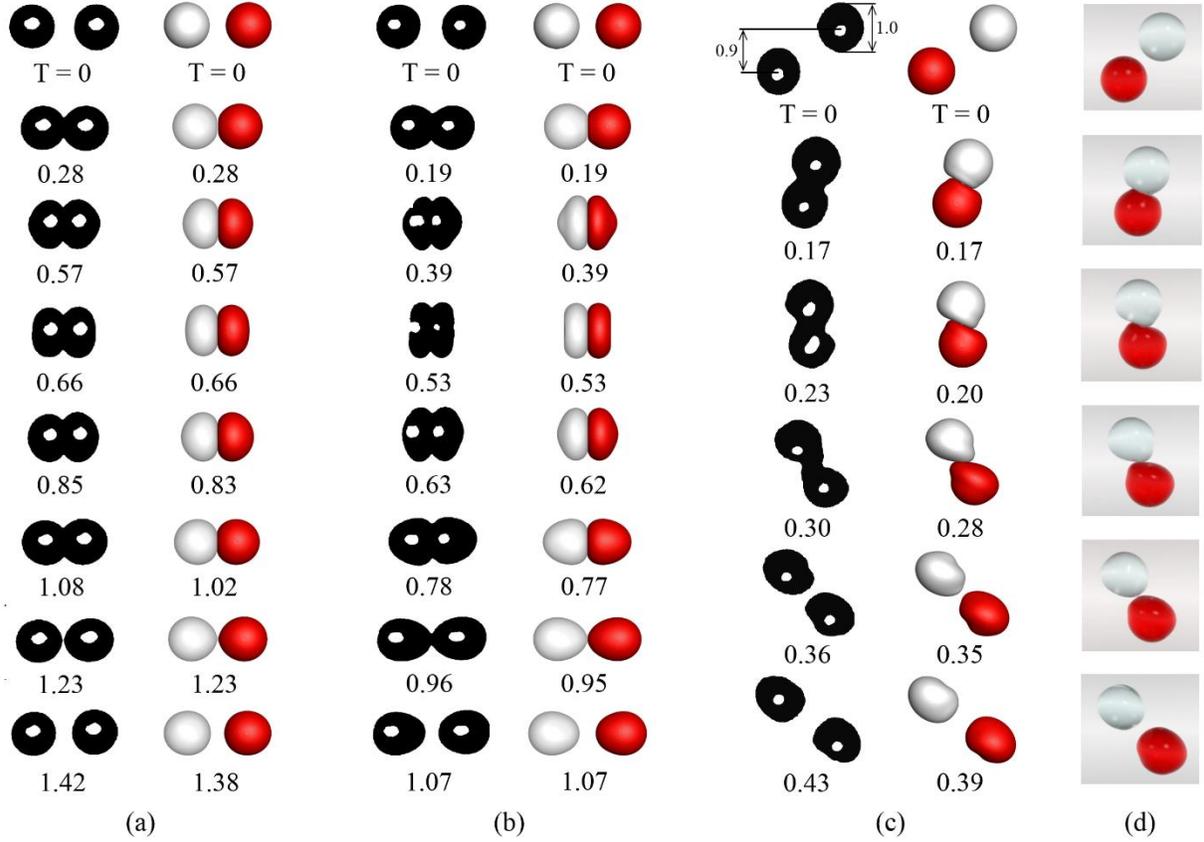

FIG. 2. Comparison between the experimental images (left) and the simulation results (right) for bouncing droplets of equal size. (a) $We = 2.3, Oh = 2.80 \times 10^{-2}, B = 0.0$ and $t_{osc} = 1.06ms$, (b) $We = 9.3, Oh = 2.78 \times 10^{-2}, B = 0.0$ and $t_{osc} = 1.03ms$, and (c) $We = 48.8, Oh = 2.80 \times 10^{-2}, B = 0.82$ and $t_{osc} = 1.06ms$. The physical time $t$ is related to the computational time T by $T = t/t_{osc}$. Additionally, the simulation results at (d) $We = 48.8, Oh = 2.93 \times 10^{-2}$ and $B = 0.9$ are adapted from Chen and Yang [45].

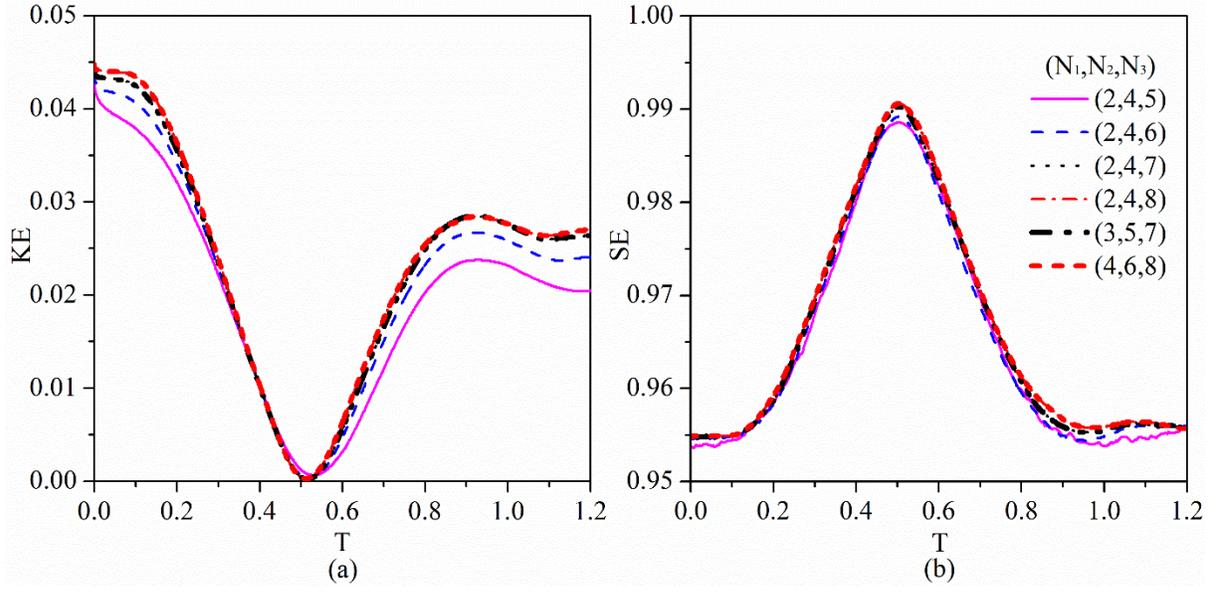

FIG. 3. Grid-independence analysis in terms of (a) the droplet kinetic energy (KE) and (b) the surface energy (SE), which is normalized by the initial total energy, for the validation case (a) shown in Fig. 2. $(N_1, N_2, N_3)$ is used to describe the mesh refinement levels in each zone of the gas, the liquid, and the droplet interface.

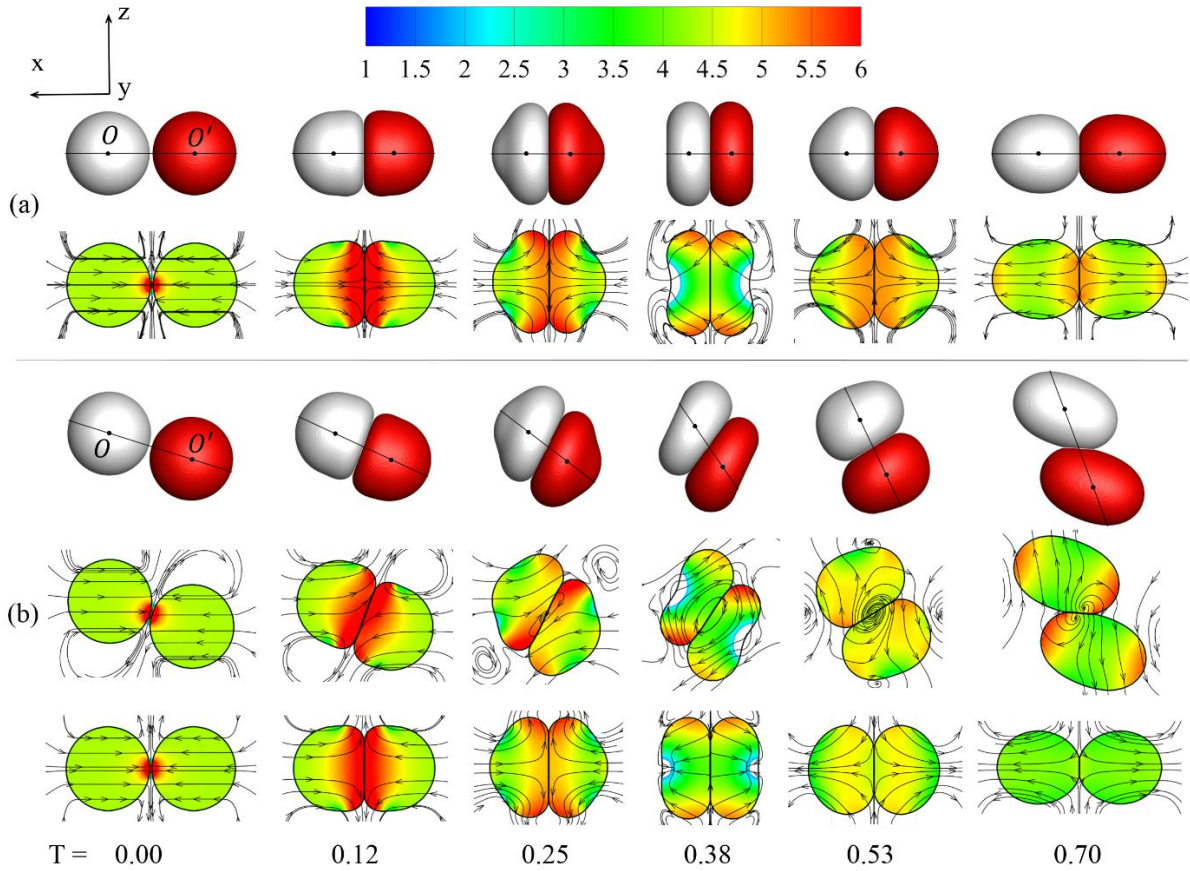

FIG. 4. Comparison of droplet deformation, pressure profiles and streamlines between (a) head-on and (b) off-center bouncing on the symmetry plane (*x-z*). For off-center bouncing, the results are also shown on the plane ($YOO'$) consisting of *y*-axis and mass center connection line.

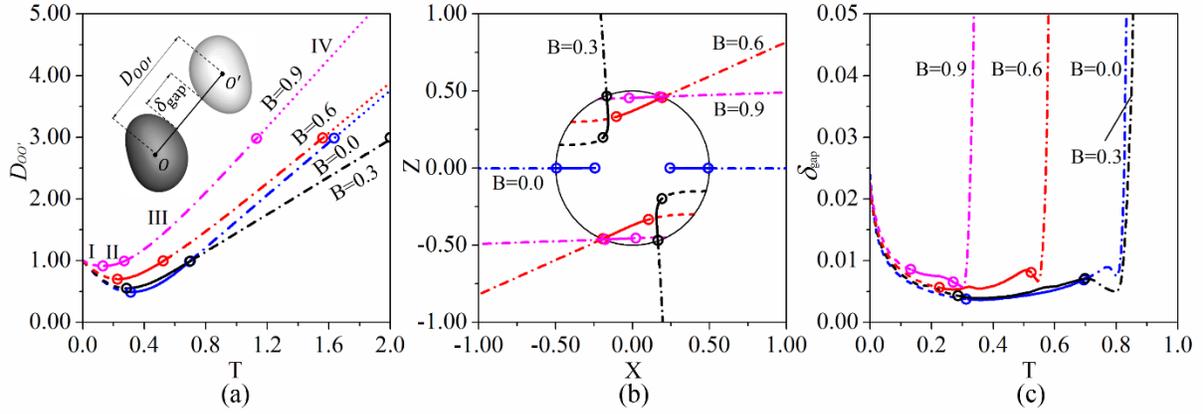

FIG. 5. Evolution of (a) the mass center distance, (b) the mass center trajectory, and (c) the gap distance for off-center droplet bouncing at $We = 9.3$ and $Oh = 2.80 \times 10^{-2}$ at different $B$s. Different stages (Stage I, II, III, and IV) are denoted by different line types (dash, solid, dash dot, and dot, respectively) and connected by open circles.

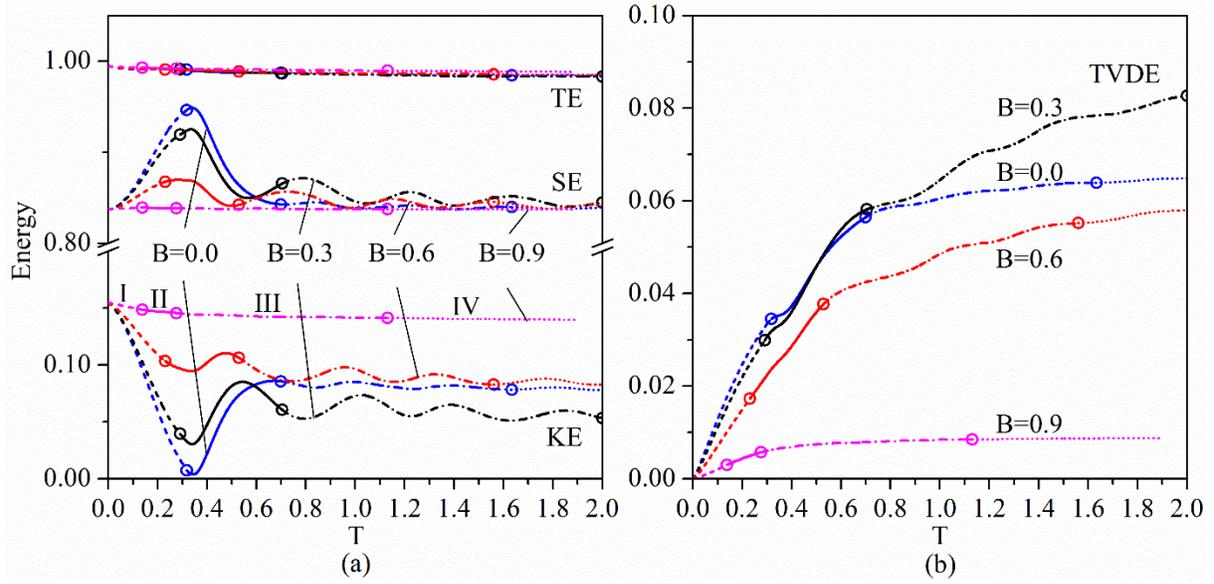

FIG. 6. Energy budget analysis of the impact parameter effects on the off-center droplet bouncing for the representative case at $We = 9.3$ and $Oh = 2.80 \times 10^{-2}$. (a) the total energy (TE), the surface energy (SE), and the kinetic energy (KE), and (b) the total viscous dissipation energy (TVDE) are of the liquid droplets and normalized by the initial total energy.

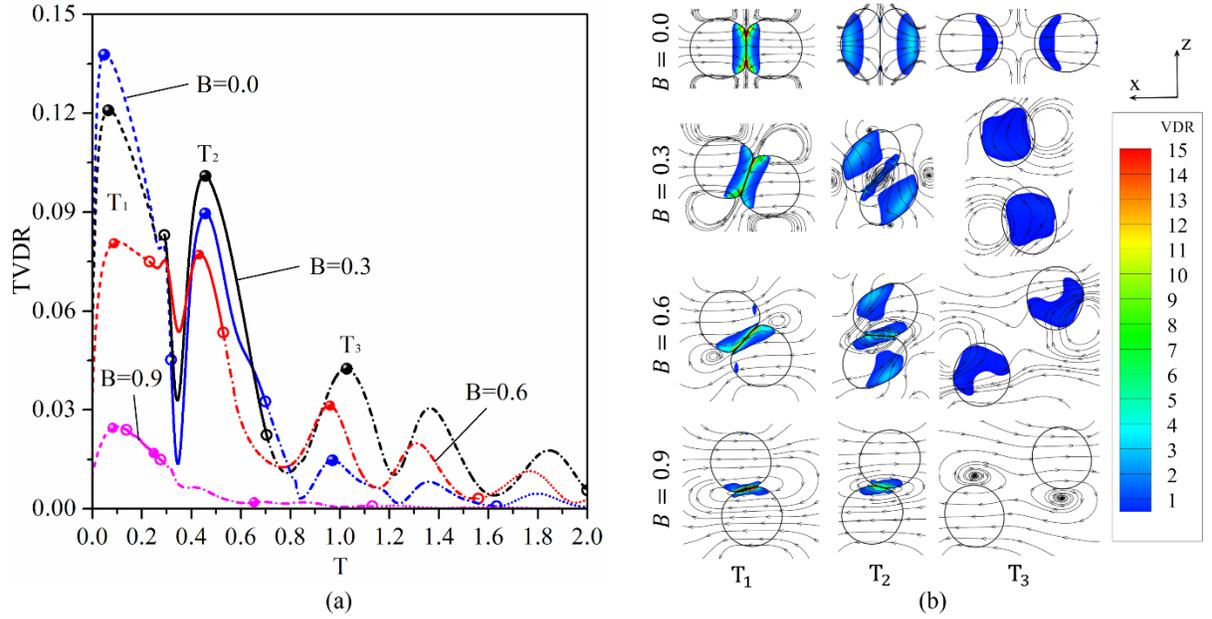

FIG. 7. Comparison of (a) the total viscous dissipation rate (TVDR) and (b) the local viscous dissipation rate (VDR) for the representative case at $We = 9.3$ and $Oh = 2.80 \times 10^{-2}$. The contours have been blanked with a low threshold value of 0.5 for clear comparison of the VDR concentration.

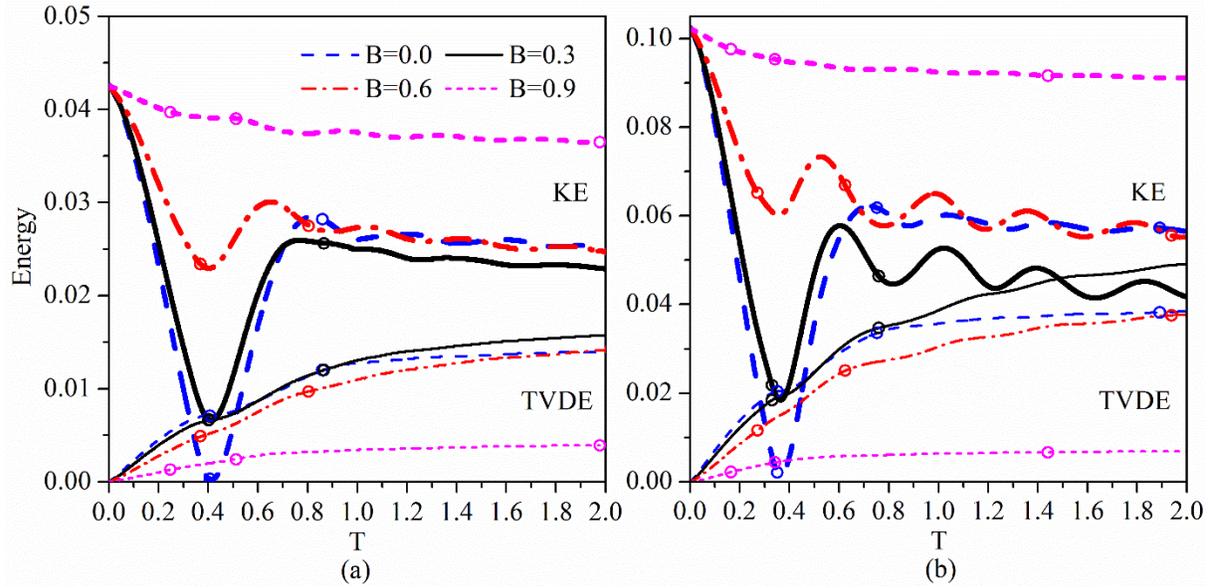

FIG. 8. Impact parameter effects on the droplets kinetic energy (KE) and the total viscous dissipation energy (TVDE) at (a) $We = 2.3, Oh = 2.80 \times 10^{-2}$ and (b) $We = 5.8, Oh = 2.80 \times 10^{-2}$.

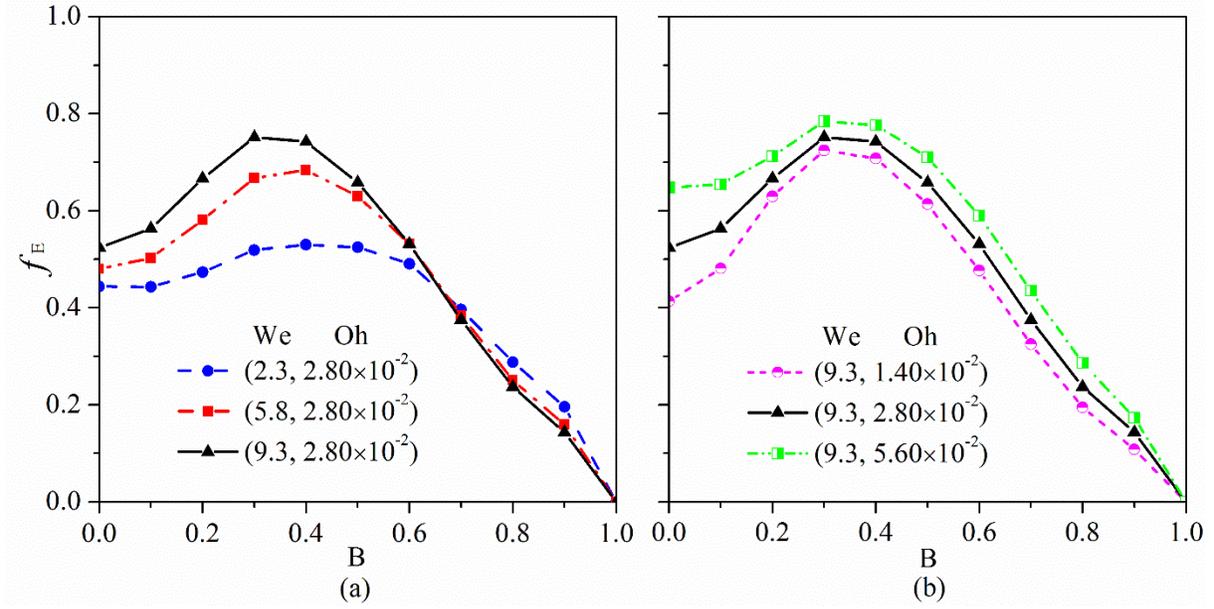

FIG. 9. Variation of the kinetic energy dissipation factor $f_E$ with impact parameters at different $We$ and $Oh$.

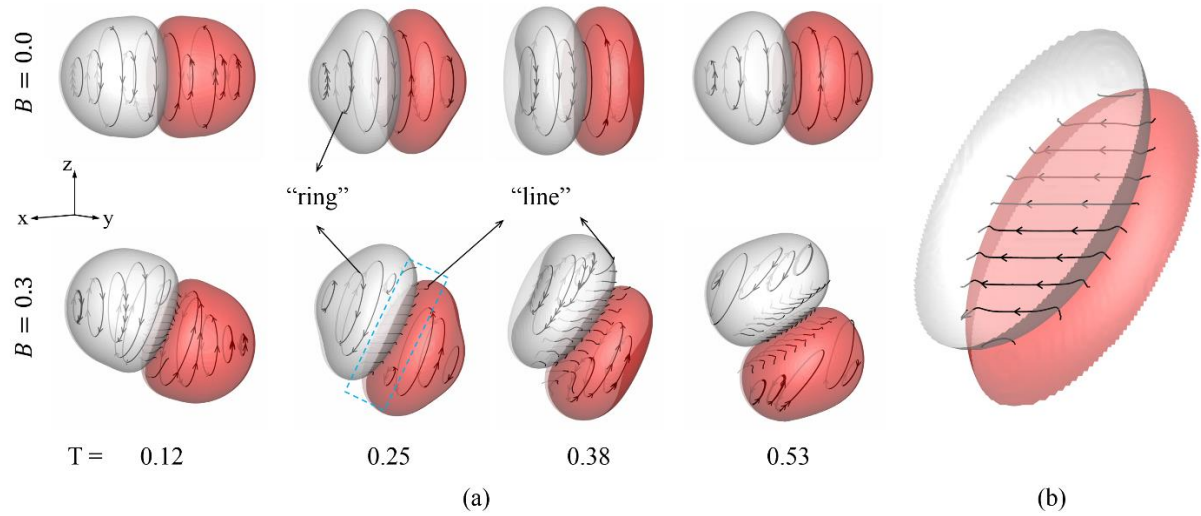

FIG. 10. Comparison of vortex lines between (a) head-on and off-center droplet bouncing and (b) a zoom-in picture of the region enclosed by the dashed frame.

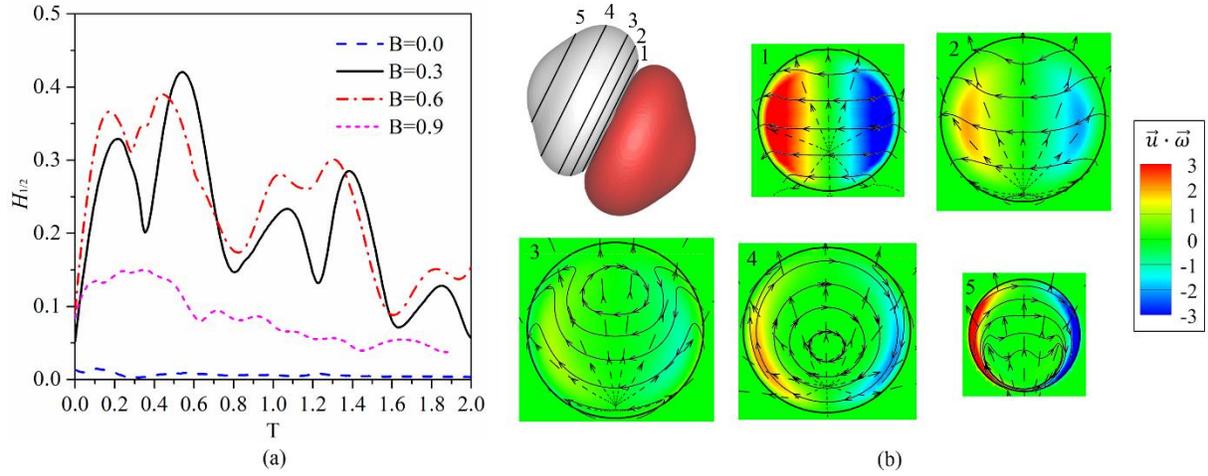

FIG. 11. Comparison of (a) a defined "helicity" between head-on and off-center droplet bouncing at different $B$s, and (b) the vortex line (solid), streamline (dash), and local distribution of $\vec{u} \cdot \vec{\omega}$ at different planes (1 to 5) by the representative time instants at T = 0.25 for $B = 0.3$.

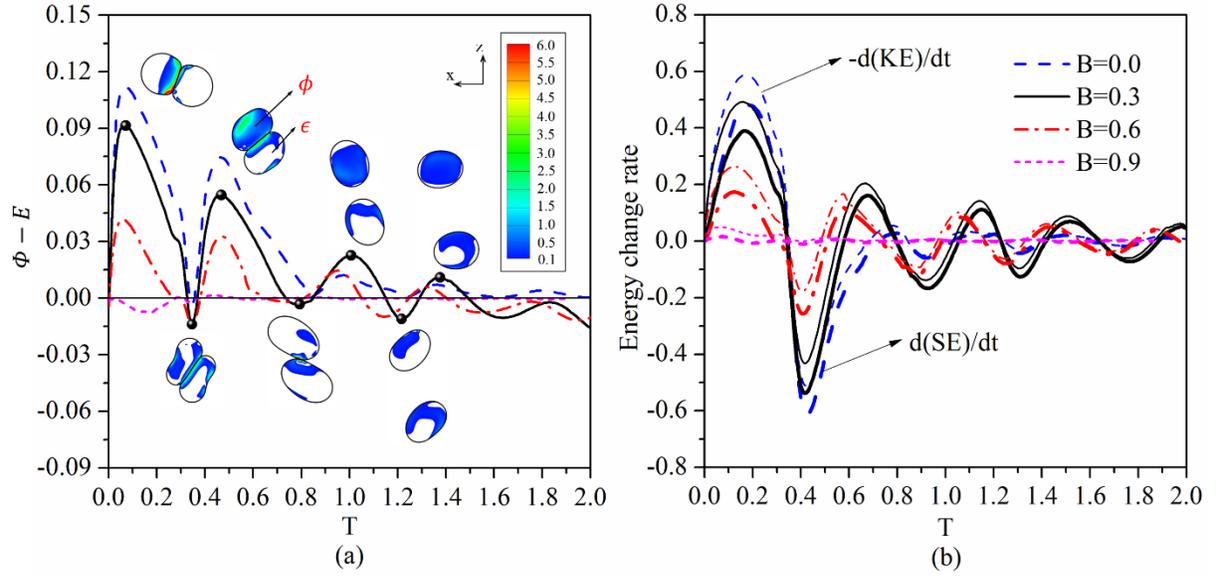

FIG. 12. Comparison of (a) the total viscous dissipation and enstrophy, and (b) the change rate of kinetic energy and surface energy for different $B$ s. The imbedded pictures are the local $\phi$ and $\epsilon$ for the representative case of $B = 0.3$ at each local extrema.